\documentclass[aps,prd,amsfonts,amssymb,preprint,eqsecnum,nofootinbib,superscriptaddress]
{revtex4} 
\usepackage{graphics,epsfig}
\def\ktr{{\tilde{k}_{tr}}}
\newcommand{\pslash}{p\llap{/\kern-0.3pt}}
\newcommand{\qslash}{q\llap{/\kern-0.3pt}}
\newcommand{\rslash}{r\llap{/\kern-0.3pt}}
\newcommand{\lslash}{\ell\llap{/\kern-0.3pt}}
\begin{document}
\preprint{WM-06-108}
%
\title{\vspace*{0.5in}  New bounds on isotropic Lorentz violation
\vskip 0.1in}
\author{Christopher D. Carone}\email[]{carone@physics.wm.edu}
\author{Marc Sher}\email[]{sher@physics.wm.edu}
\affiliation{Particle Theory Group, Department of Physics,
College of William and Mary, Williamsburg, VA 23187-8795}
\author{Marc Vanderhaeghen}\email[]{marcvdh@jlab.org}
\affiliation{Particle Theory Group, Department of Physics,
College of William and Mary, Williamsburg, VA 23187-8795}
\affiliation{Theory Center, Jefferson Lab,
12000 Jefferson Avenue, Newport News, VA 23606}
\date{September 2006}
\begin{abstract}
Violations of Lorentz invariance that appear via operators of dimension four or less are
completely parameterized in the Standard Model Extension (SME).  In the pure photonic sector
of the SME, there are nineteen dimensionless, Lorentz-violating parameters.  Eighteen of
these have experimental upper bounds ranging between $10^{-11}$ and $10^{-32}$; the
remaining parameter, $\ktr$, is isotropic and has a much weaker bound of order $10^{-4}$.
In this Brief Report, we point out that $\ktr$ gives a significant contribution to the anomalous
magnetic moment of the electron and find a new upper bound of order $10^{-8}$.  With reasonable
assumptions, we further show that this bound may be improved to $10^{-14}$ by considering the
renormalization of other Lorentz-violating parameters that are more tightly constrained. Using similar
renormalization arguments, we also estimate bounds on Lorentz violating parameters in the
pure gluonic sector of QCD.
\end{abstract}
\maketitle

\section{Introduction} \label{sec:intro}

Motivated by the extreme sensitivity of various experimental searches
for Lorentz violation, and the possibility that such searches may
provide a window into Planck-scale physics, Colladay and Kostelecky~\cite{ck}
developed the Standard Model Extension (SME) to serve as a framework for studying
Lorentz and CPT violation.  This effective theory includes all possible
Lorentz-violating terms that preserve the $SU(3)\times SU(2)\times U(1)$ gauge symmetry of
the Standard Model, have mass dimension four or less, and have coefficients that are
independent of position.  Most analyses of Lorentz violation in recent years have adopted
the parameterization of the SME.  Although Ref.~\cite{ck} considered both CPT-even
and CPT-odd terms, we will focus in this Brief Report on CPT-even terms only.

In the purely photonic sector of the SME, the CPT-even part of the
Lagrangian is
\begin{equation}
    {\cal L}=-{1\over 4}F_{\mu\nu}F^{\mu\nu} -{1\over
    4}(k_{F})_{\rho\sigma\mu\nu}F^{\rho\sigma}F^{\mu\nu} \,\,\, ,
\label{eq:lvlag}
\end{equation}
where $F_{\mu\nu}=\partial_{\mu}A_{\nu}-\partial_{\nu}A_{\mu}$.  The
coefficient $(k_{F})_{\rho\sigma\mu\nu}$ has the symmetries of the
Riemann tensor and satisfies a double-trace condition
${(k_{F})^{\mu\nu}}_{\mu\nu}=0$, leading to $19$ independent coefficients.

For ease of comparison with experiment, these $19$ coefficients are typically
rewritten in terms of four traceless $3\times 3$ matrices and a single
coefficient~\cite{mewes}.  Two of the matrices, $\tilde{k}_{e^{+}}$ and
$\tilde{k}_{e^{-}}$, have five components each and are parity even, while
$\tilde{k}_{o^{-}}$ and $\tilde{k}_{o^{+}}$ have five and three
components, respectively, and are parity odd. The remaining coefficient,
$\tilde{k}_{tr}$ is given by
\begin{equation}
\tilde{k}_{tr}={2\over 3}{(k_{F})^{j}}_{0j0}  \,\,\,.
\end{equation}
This scalar coefficient is of particular interest because it is the only one that survives
in the purely photonic sector if a requirement of spatial isotropy is imposed.  One could
easily imagine such a requirement arising due to the structure of the underlying theory.

In this context, it is particularly interesting that the experimental bound on
$\tilde{k}_{tr}$ that is quoted in the literature is substantially weaker than the bounds
on the other $\tilde{k}$ coefficients~\cite{bluhm}. The coefficients $\tilde{k}_{e^{+}}$
and $\tilde{k}_{o^{-}}$ lead to the birefringence of light in vacuo, and spectropolarimetry
of light from distant galaxies leads to bounds of order $10^{-32}$~\cite{mewes}.   Bounds
on  seven of the $\tilde{k}_{e^{-}}$ and $\tilde{k}_{o^{+}}$ coefficients have been
obtained from optical and microwave cavity experiments and are ${\cal O}(10^{-11})$ for
the $\tilde{k}_{o^{+}}$ and ${\cal O}(10^{-15})$ for $\tilde{k}_{e^{-}}$~\cite{lipa}; the
remaining coefficient in $\tilde{k}_{e^{-}}$ has been bounded by $10^{-14}$
recently~\cite{antonini}.  In contrast, the scalar coefficient, $\ktr$, has only been
bounded by approximately $10^{-4}$ from Ives-Stillwell experiments, although
experiments which substantially improve this bound have been proposed~\cite{ives}.

In this paper, we point out that the bound on $\tilde{k}_{tr}$ can be strengthened by
approximately four orders of magnitude if one considers the effect of the second term
in Eq.~(\ref{eq:lvlag}) on the anomalous magnetic moment of the electron.  This is a finite
radiative correction and the bound we obtain is relatively robust.  In addition, we show
that the presence of a non-vanishing $\tilde{k}_{tr}$ induces renormalization group
running of other Lorentz-violating operators.  This leads to even tighter constraints
provided that one is willing to assume that there is no unnatural fine-tuning of parameters.
We then apply this renormalization plus naturalness argument to obtain bounds for the first time
on the Lorentz-violating parameters in the purely gluonic sector of QCD.

\section{The anomalous magnetic moment}

\begin{figure}[t]
\epsfxsize 3.5 in \epsfbox{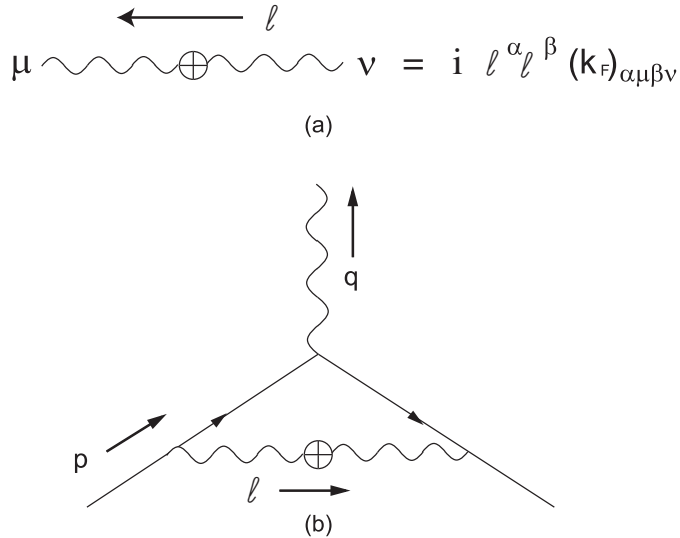} \caption{(a) Feynman rule for the Lorentz-violating
insertion. (b) Leading Lorentz-violating contribution to the electron anomalous magnetic
moment.}
\label{fig:figone}
\end{figure}

The Lagrangian of Eq.~(\ref{eq:lvlag}), together with a gauge fixing term, determines
the exact photon propagator.  From our earlier discussion, we know that the coupling $k_F$
is small.  We therefore use the conventional photon propagator in Feynman gauge and
treat the Lorentz-violating term in Eq.~(\ref{eq:lvlag}) as a perturbative insertion.
The corresponding Feynman rule is given in Fig.~(\ref{fig:figone}a).   The leading
Lorentz-violating correction to the anomalous magnetic moment of the electron then
follows from the Feynman diagram shown in Fig.~(\ref{fig:figone}b).

With momenta labelled as in the figure, the vertex correction is given by
\begin{equation}
i \Gamma^\rho(q,p) =  -e^{3}\int {d^{4} \ell \over (2\pi)^{4}}
    {[\bar{u}\gamma^{\mu}(\pslash-\lslash-\qslash+m)
    \gamma^{\rho}(\pslash-\lslash+m)\gamma^{\nu} u]\ell^\lambda \ell^{\delta}
    (k_{F})_{\lambda\mu\delta\nu}\over
    [(p-\ell-q)^{2}-m^{2}][(p-\ell)^{2}-m^{2}](\ell^{2})^{2}} \,\,\,,
\end{equation}
where $m$ is the electron mass and $u$ is its momentum space spinor wave function.
Working on shell, and extracting only the part of the amplitude with the desired
Dirac structure, one finds
\begin{equation}
i \Gamma^\rho(q,p) \supset \frac{e^3}{8\pi^2 m} k_{\mu\nu} \bar{u} [ \sigma^{\nu\rho} q^\mu -
\sigma^{\nu\alpha} q_\alpha g^{\mu\rho}] u \,\,\, ,
\label{eq:relterm}
\end{equation}
where
\begin{equation}
    k_{\mu\nu}\equiv {(k_{F})^{\rho}}_{\mu\rho\nu}.
\label{eq:kdef}
\end{equation}
With this definition, $\tilde{k}_{tr}={2\over 3}k_{00}$, and the double tracelessness
of $k_F$ implies that $k_{00}-k_{ii}=0$.

One can interpret Eq.~(\ref{eq:relterm}) as following from a simple effective Lagrangian
\begin{equation}
    {\cal L}_{eff}={1\over 8\pi^{2}}{e^{3}\over m}k_{\mu\nu}
    \overline{\psi}{\sigma^{\nu}}_{\alpha}F^{\mu\alpha}\psi  \,\,\, .
\label{eq:efflag}
\end{equation}
which can be expressed in terms of electric and magnetic fields.  Moreover, after performing a
nonrelativistic reduction of the electron spinor wave functions, one may identify the effective
quantum mechanical potential
\begin{equation}
V_{eff} = {\alpha \over \pi}(k_{00}-k_{ii})\, \vec{v}_e \times \vec{E}\cdot \vec{\mu_e}
+{\alpha\over 2\pi} \, \vec{k}\cdot\vec{E}\times\vec{\mu}_{e} + {\alpha\over \pi}k_{ij}\, \mu^{i}_e \,B^{j}
\label{eq:effpot}
\end{equation}
where $\vec{k}=k^{0i}$, $\vec{v}_e$ is the electron velocity and $\vec{\mu}_e$ is its magnetic moment operator.
The lowest-order velocity-dependent terms cancel due to the vanishing trace of $k_{\mu\nu}$.  In addition,
the bounds on the $\tilde{k}$ parameters described earlier force the off-diagonal terms in $k_{\mu\nu}$
to be extremely small. We therefore do not consider the $\vec{k}$ term in Eq.~(\ref{eq:effpot}) any further.

The fact that current bounds force all of the parameters except $\ktr$ to be extremely small implies
that $k_{00} = 3k_{11} = 3k_{22} = 3k_{33}$ is the only relevant parameter.  The coefficients $k_{ij}$ can
therefore be taken to be proportional to the identity matrix, leading to
\begin{equation}
V_{eff} = {\alpha\over 2\pi}\ktr \vec{\mu}\cdot\vec{B}.
\end{equation}
It immediately follows that there is a  Lorentz-violating contribution to the anomalous magnetic moment given by
$\Delta a \equiv (g-2)/2 = -\alpha/(2 \pi) \ktr$.  Comparing the measured anomalous magnetic moment to the most
precise QED calculations~\cite{gminus2} requires $|\Delta a| \alt 3 \times 10^{-11}$ for agreement within two
standard deviations.  We thus obtain the bound
\begin{equation}
\ktr \alt 3 \times 10^{-8}  \,\,\,.
\label{eq:bound1}
\end{equation}
This is an improvement of almost four orders of magnitude on one of the basic parameters of the SME.

\section{Renormalization Effects}

A non-vanishing value for $\ktr$ also induces renormalization group running of other Lorentz-violating
parameters.  Provided that one is willing to make relatively mild assumptions about the boundary
conditions on the Lorentz-violating couplings in the ultraviolet, one may obtain a much tighter bound
on $\ktr$ than the result presented in Eq.~(\ref{eq:bound1}). The one-loop renormalizability of
Lorentz-violating QED and the corresponding renormalization group equations have been discussed
by Kostelcky, Lane and Pickering~\cite{klp}.  Of interest to us here is the relationship between $\ktr$
and the CPT-even, Lorentz-violating parameter that modifies the electron quadratic terms,
\begin{equation}
{\cal L}_e = c_{\mu\nu}\bar{\psi}\gamma^{\mu}D^{\nu}\psi \,\,\,.
\end{equation}
In the isotropic case, the component $c_{00}$ of the coupling above and $\ktr$ are connected in a simple
way via a set of coupled renormalization group equations:
\begin{equation}
\mu{d c_{00}\over d\mu} = - \mu{d \ktr\over d\mu} = {e^{2}\over 6\pi^{2}}(2c_{00}-{3\over 2}\ktr)  \,\,\, ,
\end{equation}
If one assumes that the only nonzero SME parameter at the cutoff of the theory $\Lambda$ is
$\ktr$, then at lower energies $\mu$ one finds that the electron coupling is
\begin{equation}
c_{00}(\mu)  = \frac{3}{7} \ktr(\Lambda) \left[1-\left(\frac{\mu}{\Lambda}
\right)^\frac{7 \alpha}{3 \pi}\right]  \,\,\,,
\end{equation}
where we have ignored the running of $\alpha = e^2/4\pi$.   Choosing $\Lambda=M_{Pl}$ and
$\mu=m_e$, we would find
\begin{equation}
c_{00}(m_e) \approx 0.10 \, \ktr(M_{Pl}) \,\,\,.
\label{eq:c00me}
\end{equation}
The current bound on $c_{00}$ is $2\times 10^{-15}~$\cite{boundonc}, which leads us to conclude
\begin{equation}
\ktr \alt 2 \times 10^{-14} \,\,\,.
\label{eq:bound2}
\end{equation}
This estimate is much stronger than Eq.~(\ref{eq:bound1}), but is naive in a
number of respects. Clearly, QED with a single fermion is not the theory which describes nature all
the way up to the Planck scale, and $\ktr$ may not be the only Lorentz-violating parameter with a
non-vanishing value at the cutoff of the theory.  However, adding additional charged particle thresholds
will generically enhance the running, while the presence of other sources of Lorentz-violation in the
ultraviolet can only suppress the value of $c_{00}$ at low energies if there is an unnatural fine-tuning
of parameters. It is therefore hard to imagine in any generic theory (with a cutoff at the conventional
Planck scale) that the bound in Eq.~(\ref{eq:bound2}) can be substantially exceeded.

One can use similar renormalization and naturalness arguments to bound the $k_{F}$ parameters
of the QCD sector.  For clarity, we will call these parameters $k_{G}$. The relevant Lagrangian
is identical in form to Eq.~(\ref{eq:lvlag}), with $F_{\mu\nu}$ replaced by the non-Abelian field
strength tensor $F_{\mu\nu}^a$, and gauge indices appropriately summed.  We include an estimate of
the bounds on the $k_G$ here since there have been no published bounds on these parameters to date.
Assuming that the $k_G$ parameters are the only source of Lorentz violation at the Planck scale,
we estimate that the $c_{\mu\nu}$ coupling of a single quark field at the QCD scale is given by
\begin{equation}
c_{\mu\nu}(\Lambda_{QCD})\approx \frac{8}{9}{\alpha_{s}\over\pi}(k_{G})_{\mu\nu}(M_{Pl})
\ln(M_{Pl}/\Lambda_{QCD}),
\label{eq:srge1}
\end{equation}
where we have considered {\em only} the effect of $k_G$ on the running of $c_{\mu\nu}$.  Given our boundary
conditions, this turns out to be a reasonable approximation.  Had we done the same in our earlier QED analysis,
we would have obtained $c_{00}(m_e) \approx 0.12 \ktr(M_{Pl})$ rather than Eq.~(\ref{eq:c00me}); the difference
is irrelevant in determining an order of magnitude bound.  Since quarks are electrically charged, the
running of $c_{\mu\nu}$ generates a nonvanishing value for the tightly bounded, QED coupling $(k_F)_{\mu\nu}$.
We find that the leading effect is
\begin{equation}
(k_{F})_{\mu\nu}(\Lambda_{QCD}) \approx \frac{16}{27} N_c Q^2 {\alpha\over
\pi}{\alpha_{s}\over\pi}(k_{G})_{\mu\nu}(M_{Pl})
\ln^{2}(M_{Pl}/\Lambda_{QCD}) \,\,\,,
\label{eq:srge2}
\end{equation}
where $N_c=3$ is the number of colors and $Q$ is the quark charge in units of $e$.  Qualitatively, this
result is consistent with the statement that $(k_{F})_{\mu\nu}(\Lambda_{QCD}) \sim 0.1 (k_{G})_{\mu\nu}(M_{Pl})$.
It follows immediately that we obtain a bound on $\ktr^{QCD}$ that is of the same order as the bounds
described earlier on $\ktr$, and in particular from Eq.~(\ref{eq:bound2}),
\begin{equation}
\ktr^{QCD} \alt 2 \times 10^{-13}  \,\,\, .
\end{equation}
Alternatively, we may use the bound $c_{00}\alt 4\times 10^{-13}$~\cite{boundforc} and Eq.~(\ref{eq:srge1})
to obtain a nearly identical bound. It is important to point out, however, that the bound obtained from
$c_{00}$ might be subject to substantial improvement in the near future.  Observation of the GZK cutoff in
high-energy cosmic rays would immediately imply that $c_{00}^{p}-c_{00}^{\Delta}
\alt 2 \times 10^{-25}$~\cite{bertolami}.  Imposing a similar constraint on the quark-level couplings
would improve the bound on $\ktr^{QCD}$ by twelve orders of magnitude.

The relationship between quark and hadronic Lorentz-violating parameters deserves further comment.  From
the perspective of a low-energy effective theory for the hadrons, the presence of the $c_{\mu\nu}$ coupling
at the quark level introduces a Lorentz-violating spurion that should appear in the effective theory
describing the baryons, which are the relevant low-energy degrees of freedom.  In the previous example,
one would expect effective baryon interactions of the form
\begin{equation}
a^p c_{\mu\nu} \, \bar{p} \gamma^\mu D^\nu p + a^\Delta
c_{\mu\nu} \, \bar{\Delta}_\alpha \gamma^{\alpha\beta\mu} D^\nu \Delta_\beta + \cdots
\,\,\, ,
\end{equation}
where $\gamma^{\alpha\beta\mu}=(\gamma^\alpha\gamma^\beta\gamma^\mu-\gamma^\mu\gamma^\beta\gamma^\alpha)/2$,
$c_{\mu\nu}$ is the quark-level coupling (assuming the Lorentz-violation is isospin invariant) and
$a^p$ and $a^\Delta$ are unknown coefficients.  Naive dimensional analysis~\cite{NDA} in the effective theory
tells us that $a^p$ and $a^\Delta$ should not be much smaller than order one, since each interaction
has the same field content and derivatives as a Lorentz-conserving kinetic term.  In particular, there is
no reason in the effective theory to expect that these coefficients should be the same.  For the spatial
components of $c_{\mu\nu}$, there is additional support for this conclusion since the relationship between
quark-level and hadronic couplings can be determined reliably in a nonrelativistic quark model.   The problem
is then analogous to expressing the baryon magnetic moments as functions of the constituent quark magnetic
moments.  It is well known that the $SU(6)$ spin-flavor symmetry (for $3$ light quark flavors) of the
nonrelativistic quark model gives a good account of the octet baryon magnetic moments, and can be justified
via large-$N_c$ arguments~\cite{largen}.  In the present context, one finds for the proton and $\Delta$
(assuming three equal mass constituent quarks)~:
\begin{eqnarray}
c^{p}_{\mu \nu} &=& 4 \, c^{u}_{\mu \nu} - c^{d}_{\mu \nu} , \\
c^{\Delta}_{\mu\nu} &=& 6 \, c^{u}_{\mu\nu} + 3 c^{d}_{\mu\nu}  \,\,\,.
\end{eqnarray}
where $u$ ($d$) refer to the $u$ ($d$) quarks respectively. Assuming flavor symmetry for the Lorentz violating
coefficients, {\it i.e.} $c^{u} = c^{d} = c^{q}$, yields $c^{p}_{ij} =
3 \, c^{q}_{ij}$ and $c^{\Delta}_{ij}=3\, c^{p}_{ij}$.   Relativistic corrections can be expected to
change this estimate by around 30\%.  For the spatial parts of $c_{\mu\nu}$, the quark model calculation
is in accord with the effective field theory approach;  the $c_{\mu\nu}$ for the quarks and hadrons are
of the same order of magnitude, and that no particular cancellation between $c^{p}$ and $c^{\Delta}$ is
expected.

While we have focused on $\ktr^{QCD}$ above, other components of $(k_{G})_{\mu\nu}$ can be bounded by the
value of $(k_{F})_{\mu\nu}$ in Eq.~(\ref{eq:srge2}).  The parts of $(k_{G})_{\rho\sigma\mu\nu}$ that are
not in $(k_{G})_{\mu\nu}$ could be estimated via a two-loop calculation, which one might expect would
give a result for $(k_F)_{\rho\sigma\mu\nu}(\Lambda_{QCD})$ of order
$\alpha\alpha_s/\pi^2 \ln(M_{Pl}/\Lambda_{QCD}) (k_G)_{\rho\sigma\mu\nu}(M_{Pl})$.  We will
not pursue this in more detail since the renormalizability of the SME beyond the one-loop
level has not yet been conclusively established.

\section{Conclusions}
In this Brief Report, we have pointed out that consideration of radiative corrections in the SME
leads to a significant improvement in the bound on $\ktr$, the dimensionless, CPT-even,
Lorentz-violating parameter that appears in the photon quadratic terms and that remains in
the limit of unbroken rotational invariance.  We have shown that the Lorentz-violating
contribution to the anomalous magnetic moment of the electron leads to a bound
$\ktr \alt 10^{-8}$, while the effect of a non-vanishing $\ktr$ on the renormalization of other
Lorentz-violating operators leads to a somewhat more model-dependent bound of
$\ktr \alt 10^{-14}$.  These results provide new constraints on a variety of
Lorentz-violating models~\cite{shadmi}. We have also shown that renormalization group running leads
to similar bounds on the corresponding Lorentz-violating couplings in the gluonic sector of the SME,
parameters that have not been bounded previously in the literature. Determination of accurate
bounds on the parameters of the SME is of value given the considerable interest in
experimental tests of Lorentz violation, and in particular, in current
and planned experiments that probe the regions of the theory's parameter space that are
at present the most weakly constrained~\cite{ives}.

\begin{acknowledgments}
We thank Alan Kostelecky for useful comments.  CDC thanks the NSF for support under Grant No.~PHY-0456525.
M.S. thanks the NSF for support under Grant No.~PHY-023400.  M.V. thanks the DOE for support
under Grant No.~DE-FG02-04ER41302 and his work is  supported in part by contract DE-AC05-06OR23177 under
which Jefferson Science Associates operates the Jefferson Laboratory.

\end{acknowledgments}



\end{document}